\documentclass[aps,prd,superscriptaddress,preprint,tightenlines,nofootinbib,floatfix]{revtex4-1}

\usepackage{graphicx} % Include figure files
\usepackage{dcolumn}  % Align table columns on decimal point
\usepackage{bm}       % bold math

\usepackage{latexsym}
\usepackage{amsmath}
\usepackage{xspace}

\begin{document}

%\preprint line(s) will be ignored for PRL/PRD
%\preprint{CLEO Draft YY-NNA} % For paper draft CBX YY-NN -> Draft YY-NNA
%\preprint{CLEO CONF YY-NN}   % For conference papers
%\preprint{ICHEP ABSnnn}      % For conference papers
%\preprint{CLNS YY/NNNN}       % for CLNS notes
%\preprint{CLEO YY-NN}         % for CLNS notes

% \preprint{CPDRAFT2010-006}

%CLNS 10/2068,  CLEO 10-05
% \preprint{CLNS 10/2068}       % for CLNS notes
% \preprint{CLEO 10-05}         % for CLNS notes

% Use this form if you DO NOT have mathematical symbols in the title
% \title{Your Title Goes Here}

% Add \boldmath if you DO have mathematical symbols in the title
%\title{\boldmath Your Title Goes Here}

\title{Search for rare and forbidden decays of charm and charmed-strange mesons
to final states $\bm{h^\pm e^\mp e^+}$}

% for conference papers (ask CLEOAC for appropriate text)
%\thanks{Submitted to the 31$^{\rm st}$ International Conference on High Energy
%Physics, July 2002, Amsterdam}tmp/

%-------- INSERT HERE ------------
% Your author list goes here  REMOVE EVERYTHING to END INSERT and
% replace with your authorlist (ask cleoac).

\preprint{CLNS 10/2068}  % the CLNS number
\preprint{CLEO 10-05}    % the CLEO number

\author{P.~Rubin}
\affiliation{George Mason University, Fairfax, Virginia 22030, USA}
\author{N.~Lowrey}
\author{S.~Mehrabyan}
\author{M.~Selen}
\author{J.~Wiss}
\affiliation{University of Illinois, Urbana-Champaign, Illinois 61801, USA}
\author{J.~Libby}
\affiliation{Indian Institute of Technology Madras, Chennai, Tamil Nadu 600036, India}
\author{M.~Kornicer}
\author{R.~E.~Mitchell}
\author{M.~R.~Shepherd}
\author{C.~M.~Tarbert}
\affiliation{Indiana University, Bloomington, Indiana 47405, USA }
\author{D.~Besson}
\affiliation{University of Kansas, Lawrence, Kansas 66045, USA}
\author{T.~K.~Pedlar}
\author{J.~Xavier}
\affiliation{Luther College, Decorah, Iowa 52101, USA}
\author{D.~Cronin-Hennessy}
\author{J.~Hietala}
\author{P.~Zweber}
\affiliation{University of Minnesota, Minneapolis, Minnesota 55455, USA}
\author{S.~Dobbs}
\author{Z.~Metreveli}
\author{K.~K.~Seth}
\author{A.~Tomaradze}
\author{T.~Xiao}
\affiliation{Northwestern University, Evanston, Illinois 60208, USA}
\author{S.~Brisbane}
\author{L.~Martin}
\author{A.~Powell}
\author{P.~Spradlin}
\author{G.~Wilkinson}
\affiliation{University of Oxford, Oxford OX1 3RH, United Kingdom}
\author{H.~Mendez}
\affiliation{University of Puerto Rico, Mayaguez, Puerto Rico 00681}
\author{J.~Y.~Ge}
\author{D.~H.~Miller}
\author{I.~P.~J.~Shipsey}
\author{B.~Xin}
\affiliation{Purdue University, West Lafayette, Indiana 47907, USA}
\author{G.~S.~Adams}
\author{D.~Hu}
\author{B.~Moziak}
\author{J.~Napolitano}
\affiliation{Rensselaer Polytechnic Institute, Troy, New York 12180, USA}
\author{K.~M.~Ecklund}
\affiliation{Rice University, Houston, Texas 77005, USA}
\author{J.~Insler}
\author{H.~Muramatsu}
\author{C.~S.~Park}
\author{L.~J.~Pearson}
\author{E.~H.~Thorndike}
\author{F.~Yang}
\affiliation{University of Rochester, Rochester, New York 14627, USA}
\author{S.~Ricciardi}
\affiliation{STFC Rutherford Appleton Laboratory, Chilton, Didcot, Oxfordshire, OX11 0QX, United Kingdom}
\author{C.~Thomas}
\affiliation{University of Oxford, Oxford OX1 3RH, United Kingdom}
\affiliation{STFC Rutherford Appleton Laboratory, Chilton, Didcot, Oxfordshire, OX11 0QX, United Kingdom}
\author{M.~Artuso}
\author{S.~Blusk}
\author{R.~Mountain}
\author{T.~Skwarnicki}
\author{S.~Stone}
\author{J.~C.~Wang}
\author{L.~M.~Zhang}
\affiliation{Syracuse University, Syracuse, New York 13244, USA}
\author{G.~Bonvicini}
\author{D.~Cinabro}
\author{A.~Lincoln}
\author{M.~J.~Smith}
\author{P.~Zhou}
\author{J.~Zhu}
\affiliation{Wayne State University, Detroit, Michigan 48202, USA}
\author{P.~Naik}
\author{J.~Rademacker}
\affiliation{University of Bristol, Bristol BS8 1TL, United Kingdom}
\author{D.~M.~Asner}
\altaffiliation[Present address: ]{Pacific Northwest National Laboratory, Richland, WA 99352}
\author{K.~W.~Edwards}
\author{K.~Randrianarivony}
\author{G.~Tatishvili}
\altaffiliation[Present address: ]{Pacific Northwest National Laboratory, Richland, WA 99352}
\affiliation{Carleton University, Ottawa, Ontario, Canada K1S 5B6}
\author{R.~A.~Briere}
\author{H.~Vogel}
\affiliation{Carnegie Mellon University, Pittsburgh, Pennsylvania 15213, USA}
\author{P.~U.~E.~Onyisi}
\author{J.~L.~Rosner}
\affiliation{University of Chicago, Chicago, Illinois 60637, USA}
\author{J.~P.~Alexander}
\author{D.~G.~Cassel}
\author{S.~Das}
\author{R.~Ehrlich}
\author{L.~Fields}
\author{L.~Gibbons}
\author{S.~W.~Gray}
\author{D.~L.~Hartill}
\author{B.~K.~Heltsley}
\author{D.~L.~Kreinick}
\author{V.~E.~Kuznetsov}
\author{J.~R.~Patterson}
\author{D.~Peterson}
\author{D.~Riley}
\author{A.~Ryd}
\author{A.~J.~Sadoff}
\author{X.~Shi}
\author{W.~M.~Sun}
\affiliation{Cornell University, Ithaca, New York 14853, USA}
\author{J.~Yelton}
\affiliation{University of Florida, Gainesville, Florida 32611, USA}
\collaboration{CLEO Collaboration}
\noaffiliation

%\input{clns-10-2068.tex}
%\input{acme-short}
%\input{acme}

%-------- END INSERT ------------

%please hard code the date when you have a final draft and submit to CLEOAC
%\date{\today}
%\date{September 8, 2010}
%
\date{September 9, 2010}

\begin{abstract}
We have searched for
flavor-changing neutral current decays and lepton-number-violating decays
of $D^+$ and $D^+_s$ mesons to final states of the form $h^\pm e^\mp e^+$,
where $h$ is either $\pi$ or $K$.
We use the complete samples of \mbox{CLEO-c} open-charm data,
corresponding to integrated luminosities
of $818$ pb$^{-1}$
at the
center-of-mass energy $E_\text{CM} = 3.774$ GeV
containing $2.4 \times 10^{6}$ $D^+D^-$ pairs
and
$602$ pb$^{-1}$
at $E_\text{CM} = 4.170$ GeV
containing $0.6 \times 10^{6}$ $D^{\ast \pm}_s D^\mp_s$ pairs.
No signal is observed in any channel,
and we obtain $90\%$ confidence level upper
limits on branching fractions
$\mathcal{B}(D^{+} \rightarrow \pi^{+} e^{+} e^{-}) < 5.9 \times 10^{-6}$,
$\mathcal{B}(D^{+} \rightarrow \pi^{-} e^{+} e^{+}) < 1.1 \times 10^{-6}$,
$\mathcal{B}(D^{+} \rightarrow K^{+} e^{+} e^{-}) < 3.0 \times 10^{-6}$,
$\mathcal{B}(D^{+} \rightarrow K^{-} e^{+} e^{+}) < 3.5 \times 10^{-6}$,
$\mathcal{B}(D^{+}_{s} \rightarrow \pi^{+} e^{+} e^{-}) < 2.2 \times 10^{-5}$,
$\mathcal{B}(D^{+}_{s} \rightarrow \pi^{-} e^{+} e^{+}) < 1.8 \times 10^{-5}$,
$\mathcal{B}(D^{+}_{s} \rightarrow K^{+} e^{+} e^{-}) < 5.2 \times 10^{-5}$,
and
$\mathcal{B}(D^{+}_{s} \rightarrow K^{-} e^{+} e^{+}) < 1.7 \times 10^{-5}$.
\end{abstract}

% PACS
%11.30.Fs 	Global symmetries (e.g., baryon number, lepton number)
%11.30.Hv 	Flavor symmetries 
%12.15.Mm 	Neutral currents 
%13.20.Fc 	Decays of charmed mesons 
\maketitle
%\tableofcontents
%\listoffigures
%\listoftables

%
% Intorduction
%    physics
%    CLEO-c
%    Data sample
%
\section{\label{sec:introduction}Introduction}

As an extension of our previously reported~\cite{He:2005iz}
search for rare and forbidden decays of the $D^+$ charm meson,
$D^+ \to h^\pm e^\mp e^+$,
we report an analysis using \mbox{CLEO-c's} full open-charm data
sample for $D^+$, and also a search for $D^+_s \to h^\pm e^\mp e^+$
with CLEO-c's full $D^+_s$ data sample.
Here, $h$ is either $\pi$ or $K$, and charge-conjugate modes are
implicit throughout this article.
These decays probe
flavor-changing neutral currents (FCNC),
in $D^+ \to \pi^+ e^+ e^-$ and $D^+_s \to K^+ e^+ e^-$,
and
lepton number violations (LNV),
in $D^+ \to h^- e^+ e^+$ and $D^+_s \to h^- e^+ e^+$.
These decays are either highly suppressed or forbidden in the
standard model (SM),
but can be significantly enhanced by some non-SM physics
scenarios~\cite{Burdman:2001tf,Fajfer:2001sa,Fajfer:2005ke,Fajfer:2007dy,Ali:2001gs,Atre:2009rg}.
Standard model short-distance FCNC decays are expected to be
of order $10^{-10}$ to $10^{-9}$~\cite{Fajfer:2001sa,Fajfer:2007dy},
but long-distance vector-pole induced decays of
$D^+$ or $D^+_s$ $\to h^+ V^0 \to h^+ e^+ e^-$
(where $V^0$ is an intermediate vector meson $\rho^0$, $\omega$, or $\phi$)
are expected to be of order $10^{-6}$ to $10^{-5}$~\cite{Fajfer:2001sa,Fajfer:2007dy}.
To observe an enhancement in FCNC due to non-SM physics,
we need to search for dielectron mass regions away from the vector poles.
Measuring long-distance induced decay itself might be helpful
to understand the long-distance dynamics in the $b$ sector,
such as inclusive $b \to s \gamma$ decay
or exclusive $B \to \rho \gamma$ and $B \to K^\ast \gamma$ decays
related to extracting Cabibbo-Kobayashi-Maskawa matrix elements $|V_{t(d,s)}|$.
On the other hand,
observation of LNV ($\Delta L = 2$) decays
could be an indication of
a Majorana nature of neutrinos~\cite{Ali:2001gs,Atre:2009rg}.

%
% Data Sample
%
We have used two sets of open-charm data samples
collected by the \mbox{CLEO-c} detector in $e^+ e^-$ collisions provided by
the Cornell Electron Storage Ring (CESR).
The integrated luminosities are
$818$ pb$^{-1}$
at the
center-of-mass energy $E_\text{CM} = 3.774$ GeV
near the peak of the $\psi(3770)$ resonance which decays to
$D\bar{D}$ pairs,
and
$602$ pb$^{-1}$
at $E_\text{CM} = 4.170$ GeV near the peak of
$D^{\ast \pm}_s D^\mp_s$ pair production.
The $3.774$ GeV
data set contains $2.4 \times 10^{6}$ $D^+D^-$ pairs
and is used to study $D^+ \to h^\pm e^\mp e^+$ decays.
The $4.170$ GeV
data set
contains $0.6 \times 10^{6}$ $D^{\ast \pm}_s D^\mp_s$ pairs,
and is used to study $D^+_s \to h^\pm e^\mp e^+$ decays.

The remainder of this article is organized as follows.
The \mbox{CLEO-c} detector is described in Sec.~\ref{sec:detector}.
Event selection criteria are
described in Sec.~\ref{sec:event_selectuon}.
Features of background processes,
our suppression strategy, and signal sensitivity
are discussed in Sec.~\ref{sec:analysis}.
Results are presented as plots and tables in Sec.~\ref{sec:results}.
Systematic uncertainties associated with the branching fractions
and their upper limits are discussed in Sec.~\ref{sec:systematic_uncertainty}.
Finally, a summary of our results with systematic uncertainties
is provided in Sec.~\ref{sec:summary}.

%
% Detector
%
\section{\label{sec:detector}The CLEO-\lowercase{c} Detector}

The \mbox{CLEO-c} detector~\cite{Briere:2001rn,Kubota:1991ww,cleoiiidr,cleorich}
is a general-purpose solenoidal detector
equipped with
four concentric components:
a six-layer vertex drift chamber,
a 47-layer main drift chamber,
a ring-imaging Cherenkov (RICH) detector,
and a cesium iodide electromagnetic calorimeter,
all operating inside a 1 Tesla magnetic field provided
by a superconducting solenoidal magnet.
The detector provides acceptance of $93$\% of the full $4 \pi$
solid angle for both charged particles and photons.
The main drift chamber provides specific-ionization ($dE/dx$)
measurements that discriminate between charged pions and kaons.
The RICH detector covers approximately $80$\% of $4 \pi$ and provides
additional separation of pions and kaons at momentum above $700$ MeV.
Hadron identification
efficiencies are approximately $95$\%
with misidentification rates of a few percent~\cite{:2007zt}.
Electron identification
is based on a likelihood variable that combines
the information from the RICH detector, $dE/dx$,
and the ratio of electromagnetic shower energy to track momentum ($E/p$).
Typical electron identification efficiency is well over $90$\% on average
with the pion fake rate less than $0.1$\%
and the kaon fake rate less than a percent~\cite{Asner:2009pu,Besson:2009uv}.

A \textsc{geant}-based~\cite{geant} Monte Carlo (MC) simulation
is used to study efficiencies of signal and background events.
Physics events are generated
by \textsc{evtgen}~\cite{evtgen}, tuned with improved knowledge of
charm decays,
%~\cite{:2007zt,:2008cqa,:2009ni,pdg2008},
and
final-state radiation (FSR) is modeled by
\textsc{photos}~\cite{photos}.
Nonresonant FCNC and LNV
signal events are generated according to phase space.
%for a first order approximation.

%
% Event Selection
%
\section{\label{sec:event_selectuon}Event Selection}

Signal candidates are formed from sets of well-measured drift chamber
tracks consistent with coming from the nominal interaction point.
Charged pions and kaons are identified from the tracks
with momentum greater than $50$ MeV and with $|\cos{\theta}| < 0.93$,
where $\theta$ is the angle between the track and the beam axis.
Electron candidates are required to be above $200$ MeV
with $|\cos{\theta}| < 0.90$ to ensure that $E/p$ is well measured.

At $E_\text{CM} = 3.774$ GeV,
for each signal candidate of the form
$D^+ \to h^\pm e^\mp e^+$
(where $h$ is either $\pi$ or $K$),
two kinematic variables are computed to define a signal region:
the energy difference $\Delta E = E_{D^+} - E_\text{beam}$
and
the beam-constrained mass difference
$\Delta M_\text{bc} = [E^2_\text{beam} - {\bf p}^2_{D^+}]^{1/2} - m_{D^+}$,
where
$(E_{D^+}, {\bf p}_{D^+})$ is the four-momentum of the signal $D^+$ candidate,
$E_\text{beam}$ is the beam energy,
and
$m_{D^+}$ is the nominal~\cite{pdg2008} mass of the $D^+$ meson.
To improve the resolution of the kinematic variables,
we recover bremsstrahlung photon showers within $100$ mrad
of the direction of the electron candidates.
We define a signal box for further analysis as
$(\Delta E, \Delta M_\text{bc}) = (\pm 20 \, \text{MeV},\pm 5 \, \text{MeV})$,
which corresponds to
about $3$-standard deviations of the kinematic variables.
Because the expected contribution from the resonant decay
$\mathcal{B}(D^+ \to \phi \pi^+ \to \pi^+ e^+ e^-) \sim \mathcal{O}(10^{-6})$
is within our sensitivity,
we further subdivide $D^+ \to \pi^+ e^+ e^-$ candidates into two channels:
resonant $D^+ \to \phi (e^+ e^-) \pi^+$
and
nonresonant $D^+ \to \pi^+ e^+ e^-$ for the FCNC search.
If the dielectron invariant mass $M_{ee}$ of the signal candidate
is within $\pm 20$ MeV of the nominal~\cite{pdg2008} mass of the $\phi$ meson,
we treat it as a resonant $D^+ \to \phi (e^+ e^-) \pi^+$ candidate
and exclude it from the  $D^+ \to \pi^+ e^+ e^-$ candidates.

Similarly, at $E_\text{CM} = 4.170$ GeV,
for each signal candidate of the form $D^+_s \to h^\pm e^\mp e^+$,
the following two variables are computed to define a signal region:
the mass difference
$\Delta M = M_{D_s^+} - m_{D_s^+}$
and
the recoil mass (against the signal candidate) difference
$\Delta M_\text{recoil} (D_s^+)
= [(E_{0} - E_{D_s^+})^2 - ({\bf p}_{0} - {\bf p}_{D_s^+})^2]^{1/2} - m_{D^{\ast +}_s}$,
where $M_{D_s^+}$ is the invariant mass of the signal candidate,
$m_{D_s^+}$ is the nominal~\cite{pdg2008} mass of the $D_s^+$,
$(E_{0}, {\bf p}_{0})$ is the total four-momentum of the $e^+e^-$ beam
taking the finite beam crossing angle into account,
$(E_{D_s^+}, {\bf p}_{D_s^+})$
is the four-momentum of the signal candidate
with $E_{D_s^+} = [m^2_{D_s^+} + {\bf p}^2_{D_s^+}]^{1/2}$,
and
$m_{D^{\ast +}_s}$ is the nominal~\cite{pdg2008} mass of the $D^{\ast +}_s$.
The same bremsstrahlung recovery is performed and the
$D^+_s \to \pi^+ e^+ e^-$ channel is subdivided into resonant
$\phi(e^+e^-) \pi^+$ and nonresonant channels.
The signal box is defined as
$(\Delta M, \Delta M_\text{recoil}(D^+_s))
= (\pm 20 \, \text{MeV}, \pm 55 \, \text{MeV})$ for further analysis.
The broad recoil mass window $\pm 55$ MeV is required to allow
both primary
and secondary
(from $D^{\ast +}_s \to D^+_s \gamma$ or $D^{\ast +}_s \to D^+_s \pi^0$)
$D^+_s$ candidates
to be selected.

%
% Analysis
%
\section{\label{sec:analysis}Analysis}

Backgrounds are dominantly from events with real electrons,
particularly from $D$ semileptonic decays.  The majority of combinatorial
background events are from double charm semileptonic decays,
typically 4 or less charged particles in the event
with large missing energy due to the missing neutrinos.
Hadronic decays involving
$\gamma$-conversion and $\pi^0$ ($\eta$, $\omega$) Dalitz decay,
or accompanied by another charm semileptonic decay,
can mimic the $h^\pm e^\mp e^+$ signal, as well.
Because of the low
probability of hadrons being
misidentified as electrons~\cite{Asner:2009pu},
background from $D\bar{D}$ decays to 3-body charged-particle
hadronic decays (such as $K^- \pi^+ \pi^+$, $\pi^- \pi^+ \pi^+$, $K^0_S K^+$,
$K^+ K^- \pi^+$)
are negligible after two electrons are identified,
and they do not peak at the signal region due to the wrong mass assignments
for the hadrons misidentified as electrons.
That is,
$D\bar{D}$ backgrounds are predominantly associated with
the semileptonic decays
and
non-$D\bar{D}$ ($q\bar{q}$ continuum, $\tau$-pair, radiative return,
or QED events) backgrounds are associated with the $\gamma$-conversion
and Dalitz decays.
All of these backgrounds are nonpeaking or peak away from
the signal regions.

Our background suppression criteria tuning procedure
for $D^+ \to h^\pm e^\mp e^+$ channels
is detailed in our previous
article~\cite{He:2005iz}.
We have used the same background rejection criteria with the
four kinematic variables to reject the above-mentioned backgrounds
in $D^+$ channels
and revised the criteria to accommodate the $D^+_s$ channels.
%
% background suppression : E_other
%
The other side total energy $E_\text{other}$ is the sum of energies of
all particles other than those making up the signal candidate.
%or used in bremsstrahlung recovery.
We use this variable to reject events associated with semileptonic decays,
mainly for double charm semileptonic decays,
in which the visible other side energy would be small due to the
undetectable missing neutrinos.
We reject candidates if
$E_\text{other} < 1.0$ GeV for $D^+ \to \pi^+ e^+ e^-$,
$E_\text{other} < 1.3$ GeV for $D^+ \to K^+ e^+ e^-$,
$E_\text{other} < 1.4$ GeV for $D^+_s \to \pi^+ e^+ e^-$,
and
$E_\text{other} < 1.7$ GeV for $D^+_s \to K^+ e^+ e^-$.
For the LNV modes, we reject candidates
if the number of tracks
in the event is 4 or fewer
and
$E_\text{other} < 0.5$ GeV.
%
% background suppression : K^0_S
%
Semileptonic events involving $K^0_S \to \pi^+ \pi^-$ in the final state can mimic the signal in $\pi^+ e^+ e^-$ channels. We have used the invariant mass $M_{\pi^+\pi^-}$ to veto these events. We veto the candidate when the charged pion in the signal candidate combined with any other unused oppositely charged track satisfies $|M_{\pi^+\pi^-} - m_{K^0_S}| < 5$ MeV, where $m_{K^0_S}$ is the nominal~\cite{pdg2008} mass of the $K^0_S$.
%
% background suppression : q^2
%
Real electrons from $\gamma$-conversion and Dalitz decays are suppressed by using the dielectron invariant mass squared $q^2$ computed from the signal electron positron pair, or $q^2_\text{other}$ computed using one signal side electron (positron) combined with any oppositely charged unused track.
We veto candidates if
$q^2 < 0.01~\text{GeV}^2$
or
$q^2_\text{other} < 0.0025~\text{GeV}^2$.
%
% Mrecoil (Tag + gamma)
%
For $D^+_s$, we have required 
the solo photon from $D^{\ast +}_s$ decays to $D_s^+ \gamma$
to be explicitly reconstructed
to further suppress underlying nonstrange-charmed meson backgrounds
at $E_\text{CM} = 4.170$ GeV,
by requiring the recoil mass of the signal candidate plus solo photon
$M_\text{recoil}(D_s^+ + \gamma)$ to be within $\pm 30$ MeV of
the nominal~\cite{pdg2008} $D_s^+$ mass.
Regardless of whether the signal $D^+_s$ candidate is the primary or
secondary $D^+_s$, for the decay
$e^+ e^- \to D^{\ast \pm}_s D^\mp_s \to (D^\pm_s \gamma) D^\mp_s$,
the mass of the system recoiling against the $D^+_s$ plus $\gamma$
should peak at the $D^+_s$ mass.

The analysis was done in a blind fashion.
Before we opened the signal box,
all above-mentioned criteria
were optimized using MC events with a sensitivity
variable which is
defined as the average upper limit one
would get from an ensemble of experiments with the expected background
and no signal,
\begin{equation}
\mathcal{S} =
\frac{\sum_{N_\text{obs} = 0}^{\infty} \mathcal{C}(N_\text{obs}|N_\text{exp}) \mathcal{P}(N_\text{obs}|N_\text{exp})}
     {
       N
       \epsilon
       }
,
\label{eq:sensitivity}
\end{equation}
where
$N_\text{exp}$ is the expected number of background events,
$N_\text{obs}$ is the observed number of events,
$\mathcal{C}$ is the 90\% confidence coefficient upper limit on the signal,
$\mathcal{P}$ is the Poisson probability,
$N$ is the number of $D^+$ or $D^+_s$,
and $\epsilon$ is the signal efficiency.
In addition to the signal MC samples,
four types of background MC samples are utilized to optimize the background
suppression criteria:
$20$ times the data sample for open-charm
($D\bar{D}$, $D^\ast \bar{D}$, $D^\ast \bar{D}^\ast$, $D^\ast \bar{D}\pi$
$D^+_s D^-_s$, and $D^{\ast +}_s D^-_s$),
$5$ times the data sample of noncharm
$uds$ continuum ($q\bar{q}$),
$\tau$-pair,
and
radiative return to the $\psi(2S)$.
To normalize background
MC events to match the expected number of the data events,
we have used integrated luminosity and cross sections for each process.
For $D^+ \to h^\pm e^\mp e^+$ events at $E_\text{CM} = 3774$ MeV,
we have used
%$\sigma_{D^+D^-} = (2.91 \pm 0.03 \pm 0.05)$ nb~\cite{:2007zt},
%$\sigma_{D^0\bar{D}^0} = (3.66 \pm 0.03 \pm 0.06)$ nb~\cite{:2007zt},
$\sigma_{D^+D^-} = 2.91$ nb~\cite{:2007zt},
$\sigma_{D^0\bar{D}^0} = 3.66$ nb~\cite{:2007zt},
$\sigma_{q\bar{q}} = 13.9$ nb~\cite{Bai:2001ct},
$\sigma_{\tau^+\tau^-} = 3.0$ nb~\footnote{
With the lowest-order QED calculation,
$
\sigma(e^+e^- \to \tau^+\tau^-)
= 2 \pi \alpha^2
  \beta (3- \beta^2) / (3 s)
$,
where
$\beta = (1 - 4 m^2_\tau/s)^{1/2}$
is the $\tau$ velocity.
},
and
radiative return to the $\psi(2S)$
$\sigma_\text{RR} = 3.4$ nb~\cite{Benayoun:1999hm}.
For $D^+_s \to h^\pm e^\mp e^+$ events at $E_\text{CM} = 4170$ MeV,
we have used
%$\sigma_{D^\ast_s D_s^+} = (916 \pm 11 \pm 49)$ pb~\cite{CroninHennessy:2008yi}
%
$\sigma_{D^{\ast \pm}_s D_s^\mp} = 0.916$ nb~\cite{CroninHennessy:2008yi}
(and used other open-charm cross sections from the same reference),
$\sigma_{q\bar{q}} = 11.4$ nb~\cite{Bai:2001ct},
$\sigma_{\tau^+\tau^-} = 3.6$ nb,
and
radiative return to the $\psi(2S)$
$\sigma_\text{RR} = 0.50$ nb~\cite{Benayoun:1999hm}.
We have found that
the agreements between data and MC simulated events are excellent
in various kinematic variables used in the background suppression, giving
us confidence in our optimization procedure using our MC samples.  Possible
systematic uncertainties due to the data and MC differences are assessed
in Sec.~\ref{sec:systematic_uncertainty}.

%
% Results
%
\section{\label{sec:results}Results}

\begin{figure*}
\centering
\includegraphics[width=\textwidth]{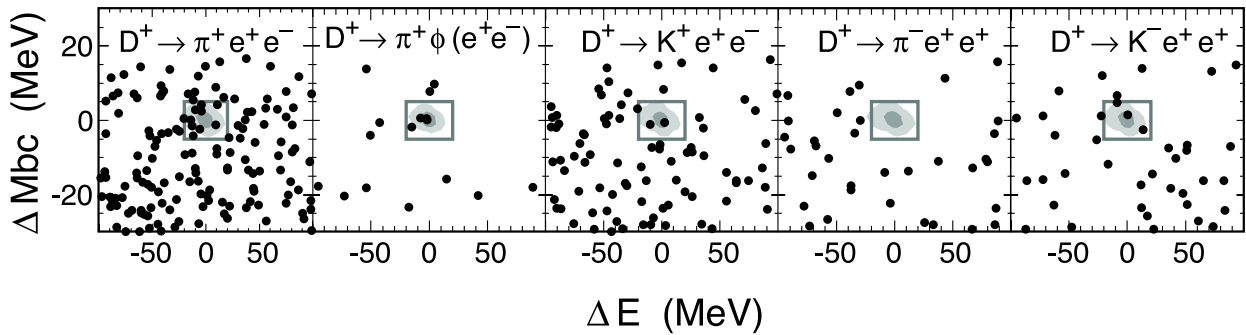}
\caption{Scatterplots of $\Delta M_\text{bc}$ vs $\Delta E$.  The two contours for each mode enclose regions determined with signal MC simulation to contain $50\%$ and $85\%$ of signal events, respectively.
The signal region, defined by
$(\Delta E, \Delta M_\text{bc}) = (\pm 20 \, \text{MeV}, \pm 5 \, \text{MeV})$,
is shown as a box.
}
\label{fig:2d-de-dmbc}
\end{figure*}

\begin{figure*}
\centering
\includegraphics[width=\textwidth]{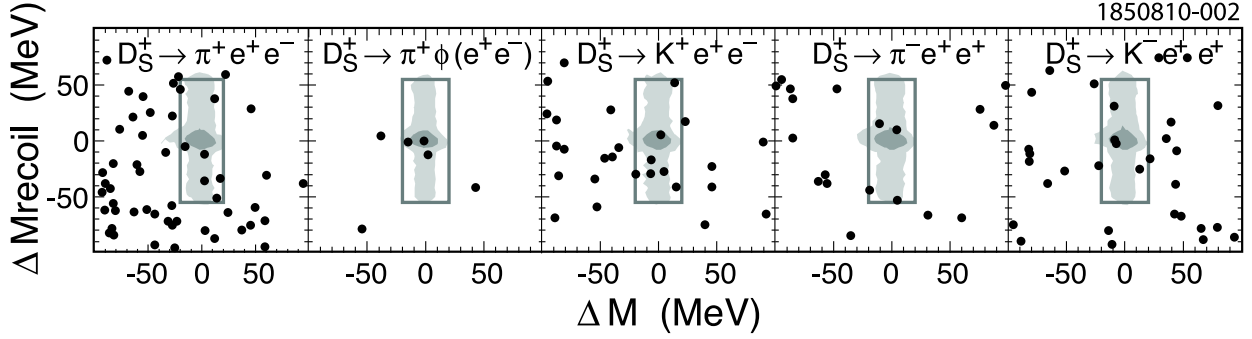}
\caption{Scatterplots of $\Delta M_\text{recoil}$ vs $\Delta M$.  The two contours for each mode enclose regions determined with signal MC simulation to contain $40\%$ and $85\%$ of signal events, respectively.
The signal region, defined by
$(\Delta M, \Delta M_\text{recoil})
= (\pm 20 \, \text{MeV}, \pm 55 \, \text{MeV})$,
is shown as a box.
}
\label{fig:2d-dm-dmrecoil}
\end{figure*}

Scatterplots of
$\Delta E$ vs  $\Delta M_\text{bc}$
and
$\Delta M(D_s^+)$ vs  $\Delta M_\text{recoil}(D_s^+)$
for signal candidates with all background suppressions applied
are shown in
Figs.~\ref{fig:2d-de-dmbc} and \ref{fig:2d-dm-dmrecoil}.
Except for the $\phi(e^+e^-) \pi^+$ channels, we find no evidence of signals,
and we calculate $90$\% confidence level upper limits (UL) on the
branching fractions
based on
\textit{Poisson processes with background}~\cite{pdg1996}
(e.g.\ Section~28.6.4 \textit{Poisson processes with background} therein)
as summarized in Table~\ref{table:summary-final-2}:
\begin{equation}
\text{UL}
=
\frac
{\mathcal{C}(N_\text{obs} | N_\text{exp})}
{N \epsilon}.\label{eq:ul}
\end{equation}
For $D^+$ and $D^+_s$ $\to \phi (e^+ e^-) \pi^+$ channels,
we find weak evidence of signals with significance
$3.5$ for the $D^+$ and $1.8$ for the $D^+_s$,
so both branching fractions and upper limits are shown in
Table~\ref{table:summary-final-2}.

\def\finalfigsz{2in}

%%% (2) Poisson - 1sigma
\begin{table*}
\centering
\caption{\label{table:summary-final-2}Upper limits on branching fractions
of $D^+$ and $D^+_s$ $\to h^\pm e^\mp e^+$ at the $90$\% confidence level
for a Poisson process~\cite{pdg1996},
where
$N$ is the number of $D^+$ (or $D^+_s$) produced in our data,
$\epsilon$ is the signal efficiency,
$N_\text{exp}$ is the number of expected background,
$N_\text{obs}$ is the number of signal candidates,
$\mathcal{C}(N_\text{obs}|N_\text{exp})$ is the $90$\% confidence
coefficient upper limit on the observed events given the expected background,
and $\mathcal{B}$ is the branching fraction or upper limit of the
branching fraction at $90$\% confidence level.
We increase the upper limits to account for systematic uncertainties by
decreasing the efficiency,
the number of $D^+$ (or $D^+_s$),
and the expected number of background each by 1 standard deviation.
For the $D^+$ and $D^+_s$ $\to \phi (e^+ e^-) \pi^+$ channels,
we have shown both branching fractions and upper limits.
}
\begin{ruledtabular}
\begin{tabular}{l cc cc c r}
Channel& $N$ & $\epsilon$ (\%)& $N_\text{exp}$& $N_\text{obs}$& $\mathcal{C}(N_\text{obs}|N_\text{exp})$& $\mathcal{B}$\\
\hline
$D^{+} \rightarrow \pi^{+} e^{+} e^{-}$ &
$4.76 \times 10^6$ &%$4761749$ &
$33.9$ &
$5.7$ &
$9$ &
$9.3$ &
%$< 5.91 \times 10^{-6}$ 
%
$< 5.9 \times 10^{-6}$ 
\\
$D^{+} \rightarrow \pi^{-} e^{+} e^{+}$ &
$4.76 \times 10^6$ &%$4761749$ &
$43.5$ &
$1.3$ &
$0$ &
$2.3$ &
%$< 1.14 \times 10^{-6}$ 
%
$< 1.1 \times 10^{-6}$ 
\\
$D^{+} \rightarrow K^{+} e^{+} e^{-}$ &
$4.76 \times 10^6$ &%$4761749$ &
$23.1$ &
$4.9$ &
$2$ &
$3.2$ &
%$< 3.03 \times 10^{-6}$ 
$< 3.0 \times 10^{-6}$ 
\\
$D^{+} \rightarrow K^{-} e^{+} e^{+}$ &
$4.76 \times 10^6$ &%$4761749$ &
$35.3$ &
$1.2$ &
$3$ &
$5.8$ &
%$< 3.54 \times 10^{-6}$ 
%
$< 3.5 \times 10^{-6}$ 
\\
$D^{+} \rightarrow \pi^{+} \phi (e^{+} e^{-})$ &
$4.76 \times 10^6$ &%$4761749$ &
$46.2$ &
$0.3$ &
$4$ &
%\multicolumn{1}{c}{$\cdot\cdot\cdot$}
 &
$(1.7^{+1.4}_{-0.9} \pm 0.1) \times 10^{-6}$ 
\\
 &
 &
 &
 &
 &
$7.9$ &
%$< 3.69 \times 10^{-6}$ 
%
$< 3.7 \times 10^{-6}$ 
\\
%\end{tabular}
%
%\begin{tabular}{l rr rr r r}
%Channel& $N$ & $\epsilon$ (\%)& $N_\text{exp}$& $N_\text{obs}$& $\mathcal{C}_{90\%}(N_\text{obs}|N_\text{exp})$& $\mathcal{B}$\\
\hline
$D^{+}_{s} \rightarrow \pi^{+} e^{+} e^{-}$ &
$1.10 \times 10^6$ & %$1103014$ &
$24.3$ &
$6.7$ &
$6$ &
$5.6$ &
%$< 2.22 \times 10^{-5}$ 
%
$< 2.2 \times 10^{-5}$ 
\\
$D^{+}_{s} \rightarrow \pi^{-} e^{+} e^{+}$ &
$1.10 \times 10^6$ & %$1103014$ &
$33.4$ &
$2.2$ &
$4$ &
$6.2$ &
%$< 1.80 \times 10^{-5}$ 
%
$< 1.8 \times 10^{-5}$ 
\\
$D^{+}_{s} \rightarrow K^{+} e^{+} e^{-}$ &
$1.10 \times 10^6$ & %$1103014$ &
$17.3$ &
$3.0$ &
$7$ &
$9.3$ &
%$< 5.19 \times 10^{-5}$ 
%
$< 5.2 \times 10^{-5}$ 
\\
$D^{+}_{s} \rightarrow K^{-} e^{+} e^{+}$ &
$1.10 \times 10^6$ & %$1103014$ &
$27.7$ &
$4.1$ &
$4$ &
$5.0$ &
%$< 1.73 \times 10^{-5}$ 
%
$< 1.7 \times 10^{-5}$ 
\\
$D^{+}_{s} \rightarrow \pi^{+} \phi (e^{+} e^{-})$ &
$1.10 \times 10^6$ & %$1103014$ &
$33.9$ &
$0.7$ &
$3$ &
%\multicolumn{1}{c}{$\cdot\cdot\cdot$}
 &
$(0.6^{+0.8}_{-0.4} \pm 0.1) \times 10^{-5}$ 
\\
 &
 &
 &
 &
 &
$6.2$ &
%$< 1.75 \times 10^{-5}$ 
%
$< 1.8 \times 10^{-5}$ 
\\
\end{tabular}
\end{ruledtabular}
\end{table*}

%
% Systematic uncertainty
%
\section{\label{sec:systematic_uncertainty}Systematic Uncertainties}

Possible sources of systematic uncertainty in our measurements
are summarized in Table~\ref{table:systsummary}.
Uncertainties associated with upper limits are classified into
three categories: uncertainties due to the normalization (the numbers of
$D^+$ and $D^+_s$), the signal efficiency, and the number of
expected background events.

\begingroup
\squeezetable
\begin{table*}
\centering
\caption{\label{table:systsummary}
Summary of systematic uncertainties in $D^+$ and $D^+_s$
$\to h^\pm e^\mp e^+$ decays.
Uncertainties associated with the branching fraction
can be classified as
three categories:
uncertainties due to the normalization (the numbers of $D^+$ or $D^+_s$),
the signal efficiency,
and
the number of background events.
The columns labeled $\pi^+ \phi$ refer to candidates with $\phi \to e^+ e^-$
decays.}
\begin{ruledtabular}
\begin{tabular}{l ccccc | ccccc}
%\hline
 &
\multicolumn{5}{c|}{$D^+$} &
\multicolumn{5}{c}{$D^+_s$}
\\
Source &
$\pi^+ e^+ e^-$ &
%$\pi^+ \phi(e^+e^-)$ &
%
$\pi^+ \phi$ &
$\pi^- e^+ e^+$ &
$K^+ e^+ e^-$ &
$K^- e^+ e^+$
&
$\pi^+ e^+ e^-$ &
%$\pi^+ \phi(e^+e^-)$ &
%
$\pi^+ \phi$ &
$\pi^- e^+ e^+$ &
$K^+ e^+ e^-$ &
$K^- e^+ e^+$
\\
\hline
Normalization &
  $2.2 \%$ &
  $2.2 \%$ &
  $2.2 \%$ &
  $2.2 \%$ &
  $2.2 \%$
&
  $5.6 \%$ &
  $5.6 \%$ &
  $5.6 \%$ &
  $5.6 \%$ &
  $5.6 \%$
\\
\hline
Tracking &
  $0.9 \%$ &
  $0.9 \%$ &
  $0.9 \%$ &
  $1.1 \%$ &
  $1.1 \%$
&
  $0.9 \%$ &
  $0.9 \%$ &
  $0.9 \%$ &
  $1.1 \%$ &
  $1.1 \%$
\\
PID &
  $2.0 \%$ &
  $2.0 \%$ &
  $2.0 \%$ &
  $2.0 \%$ &
  $2.0 \%$
&
  $2.0 \%$ &
  $2.0 \%$ &
  $2.0 \%$ &
  $2.0 \%$ &
  $2.0 \%$
\\
FSR &
  $1.0 \%$ &
  $1.0 \%$ &
  $1.0 \%$ &
  $1.0 \%$ &
  $1.0 \%$
&
  $1.0 \%$ &
  $1.0 \%$ &
  $1.0 \%$ &
  $1.0 \%$ &
  $1.0 \%$
\\
Background suppression &
  $5.0 \%$ &
  $4.2 \%$ &
  $1.5 \%$ &
  $9.4 \%$ &
  $1.5 \%$
&
  $5.2 \%$ &
  $4.5 \%$ &
  $2.1 \%$ &
  $9.0 \%$ &
  $2.2 \%$
\\
MC statistics &
  $0.6 \%$ & % 0.21/34.55*100 = 0.6
  $0.6 \%$ & % 0.22/47.13*100 = 0.5
  $0.5 \%$ & % 0.22/44.37*100 = 0.5
  $0.8 \%$ & % 0.19/23.58*100 = 0.8
  $0.6 \%$   % 0.21/35.99*100 = 0.6
%
%dE+dMbc+BG
%
%piee
% (0.23**2+.002**2 + 1.03**2+0.04**2 + (19.31/4)**2)**0.5 = 5.0
%piphi
% (0.23**2+.002**2 + 1.03**2+0.04**2 + (2.15/4)**2 + 4**2)**0.5 = 4.2
%piepep
% (0.23**2+.002**2 + 1.03**2+0.04**2 + (4.01/4)**2)**0.5 = 1.5
%kee
% (0.23**2+.002**2 + 1.03**2+0.04**2 + (37.22/4)**2)**0.5 = 9.4
%kepep
% (0.23**2+.002**2 + 1.03**2+0.04**2 + (4.09/4)**2)**0.5 = 1.5
%
&
  $0.8 \%$ & % 0.19/24.80*100 = 0.8
  $0.6 \%$ & % 0.21/34.59*100 = 0.6
  $0.6 \%$ & % 0.21/34.07*100 = 0.6
  $1.0 \%$ & % 0.17/17.67*100 = 1.0
  $0.7 \%$   % 0.20/28.31*100 = 0.7
%
%dM+dMr+dMR+BG
%
%piee
% (0.01**2+0.01**2 + 0.48**2+0.1**2 + 2.0**2 + (19.03/4)**2)**0.5 = 5.2
%piphi
% (0.01**2+0.01**2 + 0.48**2+0.1**2 + 2.0**2 + (2.23/4)**2 +4**2)**0.5 = 4.5
%piepep
% (0.01**2+0.01**2 + 0.48**2+0.1**2 + 2.0**2 + (1.48/4)**2)**0.5 = 2.1
%kee
% (0.01**2+0.01**2 + 0.48**2+0.1**2 + 2.0**2 + (35.1/4)**2)**0.5 = 9.0
%kepep
% (0.01**2+0.01**2 + 0.48**2+0.1**2 + 2.0**2 + (3.26/4)**2)**0.5 = 2.2
%
\\
Efficiency total &
  $5.6 \%$ &
  $4.9 \%$ &
  $2.9 \%$ &
  $9.8 \%$ &
  $3.0 \%$
%piee
% (0.9**2 + 2**2 + 1**2 + 5**2 +0.6**2)**0.5 = 5.6
%phipi
% (0.9**2 + 2**2 + 1**2 + 4.2**2 + 0.6**2)**0.5 = 4.9
%piepep
% (0.9**2 + 2**2 + 1**2 + 1.5**2 + 0.5**2)**0.5 = 2.9
%kee
% (1.1**2 + 2**2 + 1**2 + 9.4**2 + 0.8**2)**0.5 = 9.8
%kepep
% (1.1**2 + 2**2 + 1**2 + 1.5**2 + 0.6**2)**0.5 = 3.0
&
  $5.8 \%$ &
  $5.1 \%$ &
  $3.3 \%$ &
  $9.3 \%$ &
  $3.4 \%$
%piee
% (0.9**2 + 2**2 + 1**2 + 5.2**2 + 0.8**2)**0.5 = 5.8
%phipi
% (0.9**2 + 2**2 + 1**2 + 4.5**2 + 0.6**2)**0.5 = 5.1
%piepep
% (0.9**2 + 2**2 + 1**2 + 2.1**2 + 0.6**2)**0.5 = 3.3
%kee
% (1.1**2 + 2**2 + 1**2 + 9**2 + 1.0**2)**0.5 = 9.3
%kepep
% (1.1**2 + 2**2 + 1**2 + 2.2**2 + 0.7**2)**0.5 = 3.4
%
\\
\hline
Number of background &
  $12 \%$ &
  $68 \%$ &
  $20 \%$ &
  $12 \%$ &
  $25 \%$
%\hline
%Number of backgrounds &
%  $12 [27] \%$ &
%  $68 [210] \%$ &
%  $20 [62]  \%$ &
%  $12 [26] \%$ &
%  $24 [54] \%$
&
  $12 \%$ &
  $26 \%$ &
  $16 \%$ &
  $15 \%$ &
  $11 \%$
%\hline
%Number of backgrounds &
%  $12 [23] \%$ &
%  $26 [58] \%$ &
%  $16 [47] \%$ &
%  $15 [42] \%$ &
%  $11 [27] \%$
\\
\end{tabular}
\end{ruledtabular}
\end{table*}
\endgroup

%\subsection{\label{sec:syst_nd}Normalization}

Uncertainty in the number of $D^+$ ($D^+_s$)
is estimated by adding contributions from uncertainties in
integrated luminosity~\cite{:2007zt} $1.0$\%
and
the production cross section~\cite{:2007zt}
$2.0$\% ($5.5$\% for $D^+_s$~\cite{CroninHennessy:2008yi})
in quadrature.
We assign relative uncertainties of $2.2$\% to the number of $D^+$
and of $5.6$\% to the number of $D^+_s$.

There are several sources which can contribute to uncertainty in
the signal efficiency estimation, as listed in Table~\ref{table:systsummary}.
By adding contributions from
tracking~\cite{:2007zt},
particle identification (PID)~\cite{:2007zt,Asner:2009pu},
FSR~\cite{Besson:2009uv,Asner:2009pu},
background suppression,
and MC statistics
in quadrature
we found total uncertainties
in the signal efficiency
for each channel
range from $3$\% to $10$\%.

We use the number of background events estimated by the MC simulation
rather than using the sidebands in data. The MC samples, being
$5$-$20$ times larger, have higher precision.
We have evaluated possible systematic bias
caused by the
use of MC events rather than the data sideband by using
alternative background shapes,
and
by comparing the
MC predicted number to that interpolated from the data sideband.
We found no indication of systematic bias;
all deviations are adequately explained as statistical fluctuations
due to the data statistics. We conclude that our MC events reproduce the
features of the data backgrounds well.
We took the statistical uncertainty in the MC simulated number of backgrounds
as the systematic uncertainty in the expected number of background,
as summarized in Table~\ref{table:systsummary}.

%
% SUMMARY
%
\section{\label{sec:summary}Summary}

With the complete samples of \mbox{CLEO-c} open-charm data,
corresponding to integrated luminosities
of $818$ pb$^{-1}$
at $E_\text{CM} = 3.774$ GeV
containing $2.4 \times 10^{6}$ $D^+D^-$ pairs
and
$602$ pb$^{-1}$
at $E_\text{CM} = 4.170$ GeV
containing $0.6 \times 10^{6}$ $D^{\ast \pm}_s D^\mp_s$ pairs,
we have searched
for rare (FCNC) and forbidden (LNV) decays of $D^+$ and $D^+_s$ mesons
of the form $h^\pm e^\mp e^+$,
where $h^\pm$ is either a charged pion or a charged kaon.
We found no evidence of signals and set upper limits on branching fractions
at the $90$\% confidence level as summarized in
Table~\ref{table:summary-final-2}.
Systematic uncertainties in the signal efficiency,
the number of $D^+$ (or $D^+_s$) events,
and the expected number of background events
are incorporated by decreasing the numbers used for those quantities
by 1 standard deviation of the systematic uncertainty on those quantities.
These results are the most stringent limits on FCNC and LNV for the
$D^+$ and $D^+_s$ $\to h^\pm e^\mp e^+$ decays
to date and the limits in the dielectron channels are
comparable to those in the dimuon channels~\cite{pdg2008},
but are still a few orders of magnitude larger than the SM
expectation~\cite{Fajfer:2001sa,Fajfer:2007dy}
in FCNC decays.
This leaves some room for
possible enhancement~\cite{Burdman:2001tf,Fajfer:2001sa,Fajfer:2005ke,Fajfer:2007dy}
in both FCNC and LNV decays induced by non-SM physics.
We have separately measured branching fractions of
the resonant decays $D^+ \to \pi^+ \phi \to \pi^+ e^+ e^-$
and $D^+_s \to \pi^+ \phi \to \pi^+ e^+ e^-$ due to their large expected
contributions to $\pi^+ e^+ e^-$ channels.
The significance
of our measured branching fractions is poor at
$3.5$ standard deviations for $D^+$
and
$1.8$ standard deviations for $D^+_s$,
so we have also included upper limits in Table~\ref{table:summary-final-2}.
Our measured branching fractions of these decays
are consistent with the products of known world average~\cite{pdg2008}
branching fractions,
$\mathcal{B}(D^+ \to \phi \pi^+ \to e^+ e^- \pi^+)
= \mathcal{B}(D^+ \to \phi \pi^+)
  \times
  \mathcal{B}(\phi \to e^+ e^-)
=
[(6.2 \pm 0.7) \times 10^{-3}]
\times
[(2.97 \pm 0.04) \times 10^{-4}]
=
(1.8 \pm 0.2) \times 10^{-6}$
%
% 2.97E-4 * 6.2E-3
%
and
$\mathcal{B}(D^+_s \to \phi \pi^+ \to e^+ e^- \pi^+)
=
\mathcal{B}(D^+_s \to \phi \pi^+)
\times
\mathcal{B}(\phi \to e^+ e^-)
=
[(4.38 \pm 0.35) \times 10^{-2}]
\times
[(2.97 \pm 0.04) \times 10^{-4}]
=
(1.3 \pm 0.1) \times 10^{-5}$.
%
% 4.38E-2 * 2.97E-4
%
% .x final.C
% root> phipi_br()
%
%  B(D+ --> phi(ee) pi^+) = (1.84 +- 0.21) * 10^{6}
%  B(Ds --> phi(ee) pi^+) = (1.30 +- 0.11) * 10^{5}
%   const Double_t br_phi_2_ee  = 2.97 / 10000;
%   const Double_t dbr_phi_2_ee = 0.04 / 10000;
%
%   const Double_t br_dp_2_phipi  = 6.2 / 1000;
%   const Double_t dbr_dp_2_phipi = 0.7 / 1000;
%
%   const Double_t br_ds_2_phipi  = 4.38 / 100;
%   const Double_t dbr_ds_2_phipi = 0.35 / 100;

% CURRENT acknowledgements go here...
% download from the CLEO website 
% https://wiki.lepp.cornell.edu/lepp/bin/view/CLEO/Private/AC/JournalAcknowledgementsCurrent
%Fri Jun  4 15:43:33 EDT 2010
\begin{acknowledgments}
We gratefully acknowledge the effort of the CESR staff
in providing us with excellent luminosity and running conditions.
D.~Cronin-Hennessy thanks the A.P.~Sloan Foundation.
This work was supported by
the National Science Foundation,
the U.S. Department of Energy,
the Natural Sciences and Engineering Research Council of Canada, and
the U.K. Science and Technology Facilities Council.
\end{acknowledgments} 
%\clearpage

\end{document}